\documentclass[sigplan, screen]{acmart}
\usepackage{graphicx} 
\usepackage{listings}
\usepackage{enumitem}
\usepackage{algorithm}
\usepackage{algpseudocode}
\usepackage{caption}
\usepackage{float}
\usepackage{multirow}
\usepackage[most]{tcolorbox}
\usepackage{framed}
\usepackage{balance}

\title{On the Effectiveness of Modular Testing in EvoSuite}
\author{Elizabeth Dinella}
\affiliation {
    \institution{Bryn Mawr College}
    \country{USA}
}

\begin{abstract}
This paper explores the effectiveness of modular randomized testing for object oriented programs in Java. Modular testing involves testing individual components of a program in isolation. Often times, for effective test generation, a series of non-target setup calls must be included to obtain high coverage of the target component. In this work, we evaluate and improve modular testing with the EvoSuite test generator. We find that due to strict restrictions that disallow calls to non-target setup methods, EvoSuite's modular testing mode is ineffective and often results in low branch coverage. We propose \textsc{emote} (Effective Modular Testing with EvoSuite): an enhancement to EvoSuite that relaxes this restriction, allowing non-target methods to be included in the test prefixes. This modification draws inspiration from developer-written fuzz drivers, which often invoke setup methods to properly initialize the state before testing the target method. To ensure meaningful test generation, we modify EvoSuite's fitness function to focus branch coverage contributions on the call chain originating from the target method. \textsc{emote} is evaluated on a subset of the SF100 benchmark, showing a 15.15\% improvement in coverage of the target methods. 
\end{abstract}



\copyrightyear{2026}
\acmYear{2026}
\setcopyright{cc}
\setcctype{by}
\acmConference[SOAP '26]{Proceedings of the 15th ACM SIGPLAN International Workshop on the State Of the Art in Program Analysis}{June 15--19, 2026}{Boulder, CO, USA}
\acmBooktitle{Proceedings of the 15th ACM SIGPLAN International Workshop on the State Of the Art in Program Analysis (SOAP '26), June 15--19, 2026, Boulder, CO, USA}
\acmDOI{10.1145/3814987.3814989}
\acmISBN{979-8-4007-2709-2/2026/06}

\begin{document}

\maketitle

\section{Introduction}

Modular testing refers to testing a component of a program in isolation. Effective tests for a module can provide documentation, find bugs, and prevent regressions. Automated random testing of C programs typically requires the developer to write a testing harness or fuzz driver~\cite{libfuzzer, fudge}. The fuzz driver takes random bytes from a testing tool and often sets up the program state before invoking the target module. For example, effective testing of a \texttt{Stack} data structure's \texttt{pop} method would require adding elements (invoking \texttt{push}) before invoking the target method \texttt{pop}. 

In this paper, we explore the effectiveness of modular randomized testing of object oriented programs in Java~\cite{randoop, jqf, evosuite}. In this work, we choose to study the regression test generator, EvoSuite~\cite{evosuite, evosuite2, evosuite3}, as its generated prefixes are a target for many recent works in specification generation~\cite{toga, editas, pregen, editir, togll}. In its default mode, EvoSuite generates tests at the class level, optimizing for branch coverage of the entire unit through an evolutionary search. Under EvoSuite's modular testing mode, (triggered with the \texttt{target\_method} flag), the evolutionary search algorithm is limited to create test prefixes including only constructors of the target type and the target method. This strict limitation prevents the algorithm from selecting setup methods, resulting in low branch coverage. 

We propose \textsc{emote} (Effective Modular Testing with EvoSuite), a modification to EvoSuite's modular testing mode which allows the algorithm to include non target class methods in the prefix. This design is informed by developer practice: developer written fuzz drivers routinely invoke setup methods to drive the program under test into meaningful states before invoking the target. We argue automated test generation should follow the same pattern. However, simply relaxing the restriction is insufficient. When the target method is a callee of other class methods, a naive implementation allows non-target calls to inadvertently cover target branches, misleading the fitness function into accepting degenerate tests. We address this with a novel modification to EvoSuite's fitness function that attributes branch coverage contributions solely to the call chain originating from the target method, ensuring the evolutionary search is guided toward genuine target method coverage. We evaluate our technique on a subset of the SF100 benchmark~\cite{SF100}, and find that coverage of target methods improves by 15.15\%. Our contributions are as follows: 
\begin{enumerate}[noitemsep]
\item The first exploration of EvoSuite's effectiveness under its modular testing mode. 
\item A fork of EvoSuite with relaxed prefix constraints and a modified fitness function that attributes branch coverage solely to the target method's call chain.
\item An evaluation of \textsc{emote} on a subset of the SF100 benchmark. 
\item Open sourced artifacts of \textsc{emote}\footnote{https://github.com/elizabethdinella/emote} and our evaluation\footnote{https://github.com/elizabethdinella/SOAP-2-artifact}.
\end{enumerate}

\section{Motivating Example}
As a motivating example, consider the \texttt{checkConsistency} method from the \texttt{Jigen} project in the SF100~\cite{SF100} benchmark. The method begins with a nullness check on the \texttt{type} and \texttt{name} fields. When EvoSuite is executed in modular testing mode for the \texttt{checkConsistency} method, the algorithm is limited to generating prefixes with only calls to the constructor and the target method itself. In this example class, the \texttt{type} and \texttt{name} fields are not set during construction, and can only be set to a non-null values by calling the corresponding modifiers. However, calling these non-target methods as setup steps is not allowed by the strict constraints of EvoSuite's modular testing algorithm. As such, the generated tests never execute \texttt{checkConsistency} with non-null \texttt{type} and \texttt{name} values and results in poor coverage (1/12 branches). We find that this pattern is quite common and causes EvoSuite's modular testing mode to be ineffective in practice. Under \textsc{emote}, which allows \texttt{setType} and \texttt{setName} to be called as setup methods, coverage improves to 12/12 branches (100\%).

\begin{figure}[H]
    \centering
\begin{lstlisting}[language=Java]
void checkConsistency() {
    if (this.type != null && this.name != null) {
        if (this.type.equals("java") && this.java == null) {
            throw new AssertionError("Java element not found");
        } else {
            for(Associate associate : this.associations) {
               associate.checkConsistency();
            }

            for(Shortcut shortcut : this.shortcuts) {
               shortcut.checkConsistency();
            }

        }
    } else {
         throw new AssertionError("Check mandatory values");
    }
}

public void setType(String type) {
    this.type = type;
}

public void setName(String name) {
    this.name = name;
}
\end{lstlisting}
    \caption{Example method from SF100's \texttt{Jigen} project. EvoSuite modular testing fails to generate a target object with a non-null type and name, resulting in a low level of coverage.}
    \label{fig:mot-ex}
\end{figure}

\section{Background}
In this section, we provide necessary background on evolutionary search algorithms for object oriented test suite generation and modular fuzz testing. Modular testing refers to executing a component of the program under test in isolation. In fuzz testing, this often necessitates the use of a testing harness or fuzz driver~\cite{fudge} which typically set up state before calling the target component~\cite{libfuzzer}.

In object oriented testing, objects of the target type must be constructed in order to invoke the target method. In this paper we argue that, in a similar vein to fuzz drivers, object oriented testers must call setup methods to drive the program under test to an interesting state. 

\subsection{Evolutionary Algorithms}

Evolutionary algorithms are a class of search techniques inspired by mechanisms in evolutionary biology~\cite{GA}. In many cases, these algorithms are effective, as they maintain diversity in the population and avoid being stuck in local optima~\cite{review-of-ga}.  A general evolutionary algorithm is shown in Algorithm~\ref{alg:ga}. EvoSuite adapts this general setup to generate test suites for a target class or method. In this section, we will provide a brief overview of EvoSuite's underlying algorithm.

\begin{algorithm}
\caption{Evolutionary Algorithm}
\label{alg:ga}
\begin{algorithmic}[1]
\State Initialize population $P$ with random solutions
\State Evaluate fitness of each individual in $P$
\While{search budget not exhausted}
    \State Select individuals from $P$ based on fitness
    \State Apply crossover to produce new offspring
    \State Apply mutation to offspring
    \State Evaluate fitness of offspring
    \State Add offspring to the population $P$ 
\EndWhile
\end{algorithmic}
\end{algorithm}

\textbf{Initial Population.} EvoSuite generates the initial populations through iterative insertions of random statements. In order to avoid generation of invalid test prefixes, EvoSuite leverages typing rules and generates random statements by selecting from a pool. This pool, called the \texttt{Test Cluster}, contains three sets: test methods, generators, and modifiers. The test methods set includes all public methods on the target class and superclasses. The generators set includes constructors and factory methods for relevant types while the modifiers set contains impure methods on the target class. EvoSuite randomly chooses from these sets to insert up to \texttt{n} statements where \texttt{n} is also randomly generated up to a bound. 

\textbf{Crossover.} EvoSuite defines crossover as generating two offspring O1 and O2, from two parent tests, P1 and P2. A random value $j$ is chosen between 0 and 1, and functions as a \textit{split point} or relative value of the number of statements to include from each parent. The first offspring, O1, will contain the first $j$ fraction of statements from P1, followed by the remaining statements from P2. Similarly, O2 will contain the first $j$ fraction of statements from P2, followed by the remaining statements from P1. Since the offspring are created from members of the population, they consist exclusively of program segments from the \texttt{Test Cluster}.

\textbf{Mutation.} EvoSuite defines mutation as randomly deleting a statement, inserting a new statement, or modifying an existing statement. Insertion follows the same procedure as inserting a random statement during generation of the initial population. Modification occurs based on the type of the statement being mutated. For non-primitive statements, mutation amounts to swapping the method, field access, or constructor with an expression of the same return type. The expressions are selected from the \texttt{Test Cluster} following the same procedure executed during generation of the initial population. 

\subsection{EvoSuite Modular Testing}
EvoSuite has a modular testing mode which enables testing on a subset of class methods, specified with the \texttt{target\_method} flag. EvoSuite functions differently in the modular setting by imposing strict rules on which program components may be selected from the test cluster. The only method which may be selected from the test method set is the target method. Generators can only be selected if they constructors of the target method's class or superclass. This excludes factory methods. Modifiers are entirely excluded from selection. These restrictions are imposed during generation of the initial population. During mutation, the pool of expressions to select from during statement or insertion or modification is also limited as described. Thus, test suites generated in this mode only consist of calls to constructors of the target type and to specified target method.
\section{Related Work}
Several works have been proposed to improve EvoSuite's fitness function in the whole class testing scenario. EvoSuite's core algorithm has evolved through several generations, from optimizing a single aggregate fitness function~\cite{earlydays}, to treating each branch as a separate objective~\cite{mosa}, and to DynaMOSA~\cite{dynamosa}, which reformulates branch coverage as a many-objective optimization problem by dynamically selecting coverage targets. Beyond core algorithm design, adaptions to the fitness function have been proposed to improve test generation across various aspects. Techniques to control test length bloat~\cite{bloat}, and branch distance calculations~\cite{branch-distance} are now incorporated into EvoSuite's default configuration. However, all of these techniques target the whole class test generation scenario and are evaluated under that setting. To the best of our knowledge, this work is the first to explore modifications to EvoSuite's fitness function and test cluster in the modular testing setting.

Beyond fitness function design, other tools take different approaches to unit-level test generation. Randoop~\cite{randoop} is another widely used automated test generation tool for Java, using feedback directed random test generation rather than evolutionary search. Randoop's default mode is to test an entire program or library, but allows for modular testing functionality using the \texttt{--methodlist} flag. Like EvoSuite, Randoop imposes strict rules on which program components can be included in the generated test. Any setup methods must be manually included in a user configured list, requiring the user to anticipate which setup methods are needed, which is precisely the problem modular test generation aims to solve. 







\section{EMOTE}
Our technique, \textsc{emote}, modifies the EvoSuite modular testing mode, by allowing EvoSuite to select non-matching target methods from the \texttt{Test Cluster} during generation and mutation. This is a meaningful departure from EvoSuite's default behavior, which restricts the search to constructors and the target method alone. This limitation leads to systematically poor branch coverage in practice. In Figure~\ref{fig:mot-ex} this amounts to allowing calls to \texttt{setType} and \texttt{setName} to properly initialize the target object's state before invoking \texttt{checkConsistency}. Our only restriction is that test prefixes must end with a call to the target method, ensuring the generated tests directly exercise the target. \textsc{emote} is implemented as a fork of EvoSuite version 1.1.0.

\subsection{Selective Coverage Contribution}
A key challenge arises when the target method is a callee of other class methods. Since our technique allows non target methods to be included in the prefix, these calls may inadvertently execute the target method as a helper, contributing to its branch coverage before the target method is directly invoked. This misleads the fitness function into accepting a test as achieving high coverage without ever meaningfully exercising the target method under direct invocation.  Such tests are undesirable in the modular testing setting, where the goal is to exercise the target method's own logic directly through its own invocation, not through indirect calls from other methods.

As a motivating example, consider a method \texttt{stripString}, from the \texttt{a4j} project in the SF100 benchmark. The method filters the characters of \texttt{input}, retaining only those that appear in \texttt{allowedSet}. A modular test suite achieving full branch coverage of \texttt{stripString} should execute the method with (1) an \texttt{input} containing characters in \texttt{allowedSet} and (2) an \texttt{input} containing characters not in the \texttt{allowedSet}. 

The surrounding class also includes a method \texttt{getPrice} which calls \texttt{stripString} with a hardcoded \texttt{allowedSet} of digits. When testing the target method \texttt{stripString},  our modification allows EvoSuite to select \texttt{getPrice} from the \texttt{Test Cluster}. Because \texttt{getPrice} internally calls the target method, this prefix inadvertently covers target branches without  \texttt{stripString} being directly invoked. This leads to the generation of \texttt{test0} shown in Figure~\ref{fig:callee-test}, which achieves 100\% branch coverage with only a single direct call to the target method, leaving the false branch of the \texttt{indexOf} check covered only through \texttt{getPrice}. This is undesirable in a modular testing scenario, where branches should be covered through direct invocation of the target method. 

The fitness function modification is not intended to increase coverage, but to ensure that reported coverage reflects genuine exercise of the target method under direct invocation. A higher coverage number achieved through indirect calls does not constitute a better modular test suite. 

\begin{figure}[htbp]
    \centering
\begin{lstlisting}[language=Java]
public String stripString(String allowedSet, String input) {
    StringBuffer result = new StringBuffer();
    for (int i = 0; i < input.length(); i++) {
        char c = input.charAt(i);
        if (allowedSet.indexOf(c) != -1) {
            result.append(c);
        }
    }
    return result.toString();
}

public BigDecimal getPrice(String priceStr) {
    String digits = ".0123456789";
    String stripped = this.stripString(digits, priceStr);
    ...
}
\end{lstlisting}
    \caption{Example method from SF100's \texttt{a4j} project. The target method \texttt{stripString} is called by \texttt{getPrice}, a method on the same class.}
    \label{fig:callee}
\end{figure}

\begin{figure}[htbp]
    \centering
\begin{lstlisting}[language=Java]
public void test0() throws Throwable {
    a4jUtil a4jUtil0 = new a4jUtil();
    a4jUtil0.stripString("k9mX", "k9mX");
    a4jUtil0.getPrice("qB#");
}
\end{lstlisting}
    \caption{An EvoSuite generated test for the \texttt{stripString} method. A call to \texttt{getPrice} should not contribute to the coverage, although it does execute the target method \texttt{stripString}.}
    \label{fig:callee-test}
\end{figure}


To address this challenge, we modify EvoSuite's fitness function such that branch coverage only counts toward an individual's fitness when the origin of the call chain is the target method. This prevents branches covered through non-target callers from artificially inflating the fitness evaluation.

\section{Evaluation}

\begin{table*}[ht]
\centering
\small
\begin{tabular}{lrrrrrrr}
\toprule
\multirow{2}{*}{Project} & \multirow{2}{*}{\# of Methods} & \multirow{2}{*}{\# of Branches} & \multicolumn{2}{c}{EvoSuite (Plain)} & \multicolumn{2}{c}{\textsc{emote}} & \multirow{2}{*}{\% Change} \\
\cmidrule(lr){4-5} \cmidrule(lr){6-7}
 & & & \# of Covered Branches & \% Coverage & \# of Covered Branches & \% Coverage & \\
\midrule
tullibee & 157 & 673 & 374.00 & 55.57\% & 388.00 &  57.65\% & +2.08\% \\
a4j & 466 & 883 & 528.33 & 59.83\% & 745.67 & 84.45\% & +24.61\% \\
jigen & 161 & 262 & 174.67 & 66.67\% & 243.33 & 92.87\% & +26.21\% \\
rif  & 32 & 62 & 32.00 & 51.61\% & 33.00 & 53.23\% & +1.61\% \\
templateit & 41 & 129 & 111.67 & 86.57\% & 115.00 & 89.15\% & +2.58\% \\
\midrule
Total & 857 & 2009 & 1220.67 & 60.76\% & 1525.00 & 75.91\% & +15.15\% \\
\bottomrule
\end{tabular}
\caption{Average modular coverage of the first five projects in the SF100 benchmark.}
\label{tab:coverage}
\end{table*}

Our evaluation investigates the effect of our modification on EvoSuite's modular testing mode. We evaluate \textsc{emote} on a subset of the SF100~\cite{SF100} benchmark. 

\begin{figure}[htbp]
    \centering
    \includegraphics[width=0.8\linewidth]{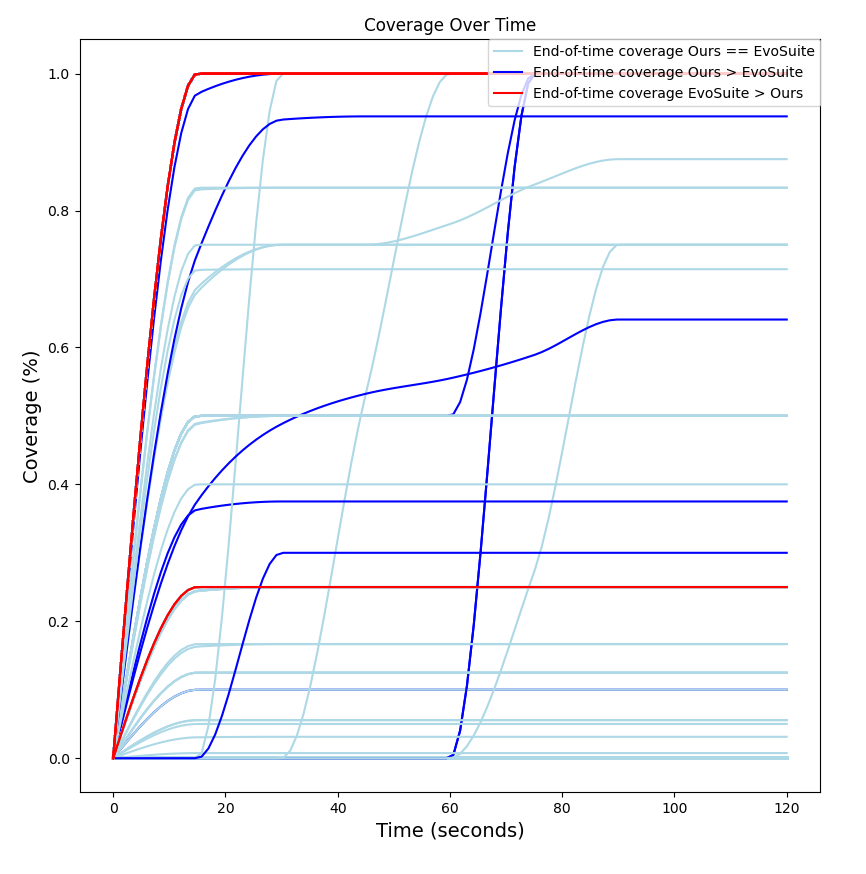}
    \caption{Average Branch Coverage Over Time.}
    \label{fig:timeline}
\end{figure}

\textbf{Experimental Setup}
We evaluate EvoSuite's modular testing mode on the first five projects of the SF100 benchmark~\cite{SF100}, which has been identified as a target for sound empirical studies of Java software testing. As a baseline, we run EvoSuite with the target method flag on each method in our benchmark.  We exclude constructors and static initialization blocks as EvoSuite does not support modular testing of initializers. For both scenarios (with and without our modification), we run EvoSuite with the following parameters: DynaMOSA~\cite{dynamosa} goal management, branch coverage criterion, and JUnit checks turned off. We set a search budget of 120 seconds for each method, consistent with standards in literature~\cite{SF100, codamosa} and validated empirically on a held-out project. Figure~\ref{fig:timeline} confirms that coverage stabilizes well before 120 seconds across methods, suggesting that a longer budget would not meaningfully change the results. As coverage improvements occur primarily in the first 20 seconds as setup calls are selected, we expect the relative improvement of our technique over the baseline to persist across different budget sizes.


In this evaluation, we answer the research question: \textit{Given the same search budget, does allowing for selection of non-target methods lead to higher branch coverage?} Table~\ref{tab:coverage} reports the end-of-search branch coverage and percent improvement of our technique on each project. Note that the total number of branches is lower than the number reported in the SF100 paper as we do not include constructors in our evaluation. \textsc{emote} achieves an overall branch coverage improvement of 15.15\%, covering 304 additional branches across the benchmark. In total, 162 methods saw coverage improvements under our modification. The highest improvement was in the \texttt{jigen} project at over 25\% where many methods follow the state-initialization pattern illustrated in Figure~\ref{fig:mot-ex}, suggesting that relaxing the prefix constraint directly enables coverage of previously unreachable branches.

The smallest improvement was on the \texttt{rif} project since many of the methods are static and thus do not require state setup on a target object. There were only 7 methods where \textsc{emote} resulted in lower branch coverage.  We discuss the reason for the lower coverage on these samples in Section~\ref{sec:eval:err}. 

\begin{figure}[htbp]
    \centering
    \includegraphics[width=0.8\linewidth]{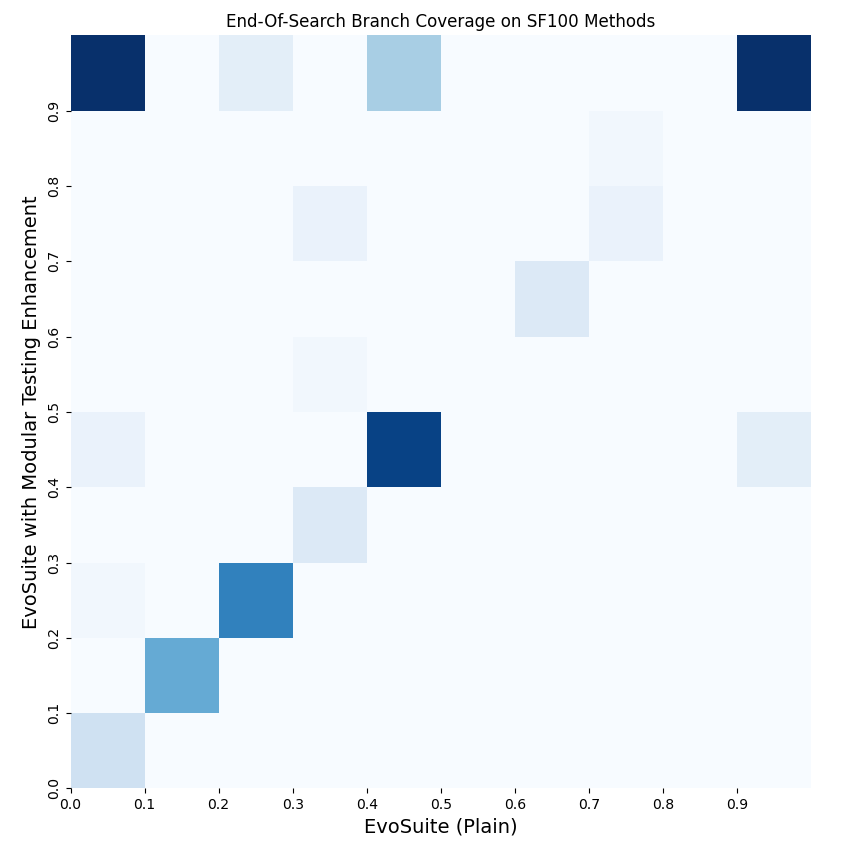}
    \caption{Heatmap of Average Branch Coverage.}
    \label{fig:plot}
\end{figure}

In Figure~\ref{fig:plot}, we plot a heat map of the end-of-search branch coverage achieved using \textsc{emote} (y axis) and the default EvoSuite coverage (x axis) for each method in our benchmark. Samples above the \texttt{x=y} line indicate that \textsc{emote} achieved higher coverage than EvoSuite's default algorithm. Here we see that many samples are concentrated in the top-left and top-right corners. The top-left indicates that EvoSuite achieved 0\% branch coverage while \textsc{emote} achieved 100\% branch coverage. The top-right indicates that both algorithms achieved full branch coverage. Most other non-zero components of the grid are above the \texttt{x=y} line, indicating that our modification was effective in increasing branch coverage. 

Lastly, we plot the branch coverage over time in Figure~\ref{fig:timeline}. We see that most coverage improvements occur in the first 20 seconds as setup calls are selected early in the search.  The majority of lines fall at or above the baseline coverage, indicating that \textsc{emote} rarely decreases coverage relative to the default EvoSuite configuration.



\begin{framed}
\noindent The evaluation showed an improvement in total branch coverage. With the same 
search budget, \textsc{emote} covered 304 additional branches. This resulted 
in a 15.15\% total improvement in branch coverage.
\end{framed}

\subsection{Error Analysis}
\label{sec:eval:err}
In our evaluation, there were 7 methods in which EvoSuite without our modification resulted in higher branch coverage. In this section we explore the reason for this performance degradation. Consider the target method \texttt{getArtist} in Figure~\ref{fig:err}. Here, the \texttt{artists} field is public and can be directly modified without calling any non-target class methods. Thus, by loosening the restriction on methods we can select, we increase the search size without advantage. The 7 methods for which \textsc{emote} results in lower performance all follow this pattern of a public field that can be set without calls to setup methods.

\begin{figure}[htbp]
    \centering
\begin{lstlisting}[language=Java, linewidth=0.96\linewidth]
public class Artists implements Serializable {
   ArrayList artists;

   public Artists() {
   }

   public String getArtist(int var1) {
      String var2 = null;
      if (this.artists.size() - 1 > var1) {
         var2 = (String)this.artists.get(var1);
      }

      return var2;
   }

   ...
}
\end{lstlisting}
    \caption{Class from the \texttt{a4j} project. Branch coverage of the target method \texttt{getArtist} is lower with \textsc{emote}.}
    \label{fig:err}
\end{figure}

\section{Conclusion}
In conclusion, EvoSuite's existing modular testing feature is ineffective in practice due to strict restrictions that prevent the inclusion of setup methods in generated test prefixes. We propose an approach to improve coverage of target methods by loosening the restriction on method calls which can be inserted during the evolutionary search. Evaluated on a subset of SF100, our approach has 15.15\% improvement in branch coverage. 

\balance
\bibliographystyle{ACM-Reference-Format}
\bibliography{references}  

\end{document}